\documentstyle[11pt,newpasp,twoside]{article}
\input{epsf}

\begin{document}

\title{Evolution of density profiles of dark matter halos}
\author{Ewa L. \L okas}
\affil{Copernicus Center, ul. Bartycka 18, 00-716 Warsaw, Poland}

\begin{abstract}
I apply the modified spherical infall model to test the dependence of the
shape of density profiles of dark matter halos on redshift. The model
predicts that the density profile of a halo of a given mass steepens
with time and its concentration parameter changes like $c \propto 1/(1+
z_{0})^2$.
\end{abstract}

%\keywords{}

\section{Introduction}
$N$-body simulations with power-law initial power spectra
(Navarro, Frenk, \& White 1997, hereafter NFW) suggest that density
profiles of dark halos in large range of masses are well fitted by a
simple universal formula that steepens from $r^{-1}$ near
the center of the halo to $r^{-3}$ at large distances. The shape of the
profile can be described by the concentration parameter
$c = r_{v}/r_{\rm s}$ where $r_{v}$ is the virial radius and $r_{\rm s}$
is the scale radius at which the slope is $-2$. The concentration
parameter was observed to decrease with the mass of the halo for a wide
range of cosmological models, that is in general the smaller the halo mass
the steeper its profile.

The improved spherical infall model (hereafter SIM) of \L okas (1999)
reproduces this behaviour in the case of $\Omega=1$ universe and
scale-free power spectra. The improved version of the model involves the
generalized form of the initial density distribution and introduces a
cut-off in this distribution at the distance equal to half the
separation between typical density peaks.

Recently Bullock et al. (1999) observed in their $N$-body simulations of
$\Lambda$CDM model that the dependence of concentration parameter of a
halo of given mass on redshift $z_{0}$, at which the halo profiles are
measured, is $c \propto 1/(1+ z_{0})$. As shown in the next section, the
same behaviour follows from the model proposed by NFW to describe their
results in the case of scale-free spectrum. My purpose here was to check
whether such dependence on redshift can be reproduced by SIM.

\section{Results}

The improved SIM can be generalized to obtain the dependence on $z_{0}$ in
a straighforward way by asuming that the collapse time of the halo is not
the present epoch, but is equal to the age of universe at $z_{0}$. Other
quantities, like the critical density, the proper virial radius or the
mass of the halo, have to be changed accordingly to incorporate the
dependence on redshift.

\begin{figure}
\begin{center}
    \leavevmode
    \epsfxsize=12cm
    \epsfbox[43 33 562 314]{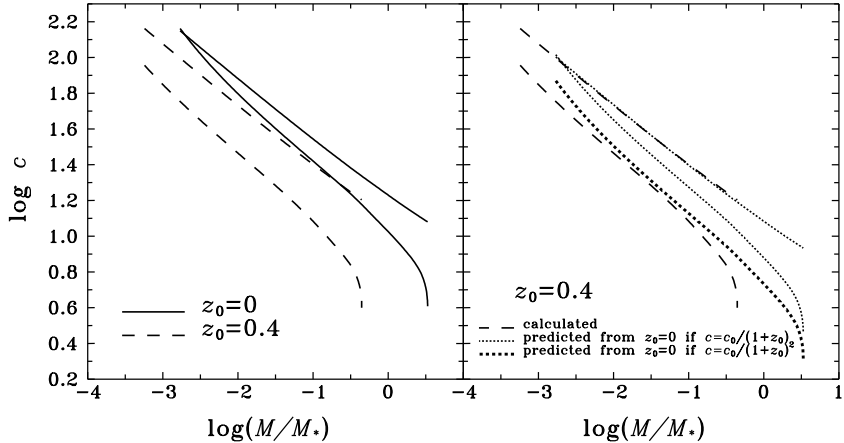}
\end{center}
    \caption{The dependence of the relation between concentration
    parameter and mass on redshift.}
\label{koncz}
\end{figure}

The left panel of Figure~\ref{koncz} shows the dependence of $c$ on the
mass of the halo expressed in units of the present nonlinear mass
$M_{\ast}$ for the present epoch, $z_{0}=0$, and at $z_{0}=0.4$. All
results were obtained for $\Omega=1$ and scale-free power spectrum of
index $n=-1$. The lines bending at larger masses show the SIM
results for initial redshifts between $z_{\rm i}=1500$ and
$z_{\rm i}=40$ while the almost straight ones correspond to the NFW
results or the toy model proposed by them to describe the dependence on
$z_{0}$. The NFW results are shown in the same range of masses as obtained
in SIM.

The right panel of the Figure repeats the $z_{0}=0.4$ results and shows
also predictions for $c$ at $z_{0}=0.4$ calculated with assumptions that
$c(z_{0}) = c_{0}/(1+z_{0})$ (thinner dotted lines) or $c(z_{0}) =
c_{0}/(1+z_{0})^2$ (thicker dotter line only for SIM results), where
$c_{0}=c(z_{0}=0)$. The Figure shows that the NFW results indeed behave as
observed by Bullock et al. (1999) (the dashed and dotted straight lines
overlap) while the SIM results are reproduced better by the
$1/(1+z_{0})^2$ behaviour of the concentration parameter. Similar
dependence on redshift is found for other values of $z_{0}$.
Therefore SIM reproduces only qualitatively the redshift dependence of
the shape of the profiles found in $N$-body simulations.

\acknowledgments
This work was supported in part by the Polish State Committee for
Scientific Research grant No. 2P03D00813.

\end{document}